\documentclass[superscriptaddress,aps,prd,twocolumn]{revtex4}
\usepackage{graphicx}
\usepackage{amsmath}
\usepackage{bm}
\usepackage{color}
\begin{document}
\title{Mode exchange and dynamical tunneling in the annular billiard}
\author{Yu.\,L.\,Bolotin}
\email{ybolotin@gmail.com}
\author{V.\,A.\,Cherkaskiy}
\email{vcherkaskiy@gmail.com}
\affiliation{A.I.Akhiezer Institute for Theoretical Physics,
National Science Center "Kharkov Institute of Physics and
Technology", Akademicheskaya Str. 1, 61108 Kharkov, Ukraine}
\author{K.\,A.\,Lukin}
\affiliation{Usikov Institute of Radiophysics and Electronics, 12, Proskura st., Kharkov, 61085, Ukraine }
\author{I.\,Yu.\,Vakulchik}
\affiliation{Kharkov Karazin National University, Svobody Sq. 4, 61077 Kharkov, Ukraine}
\date{\today}
\begin{abstract}
We present numerical results on calculations of energy spectra and wave functions of annular billiard.
\end{abstract}
\maketitle
\section{Introduction}
\label{intro}
In recent decades a new direction of investigations started its active developing devoted to studies of the so-called quantum chaos. This term implies analysis of behavior of the quantum systems whose classical counterpart demonstrates chaotic dynamics.

\begin{figure}
\begin{center}
\includegraphics[width=\columnwidth]{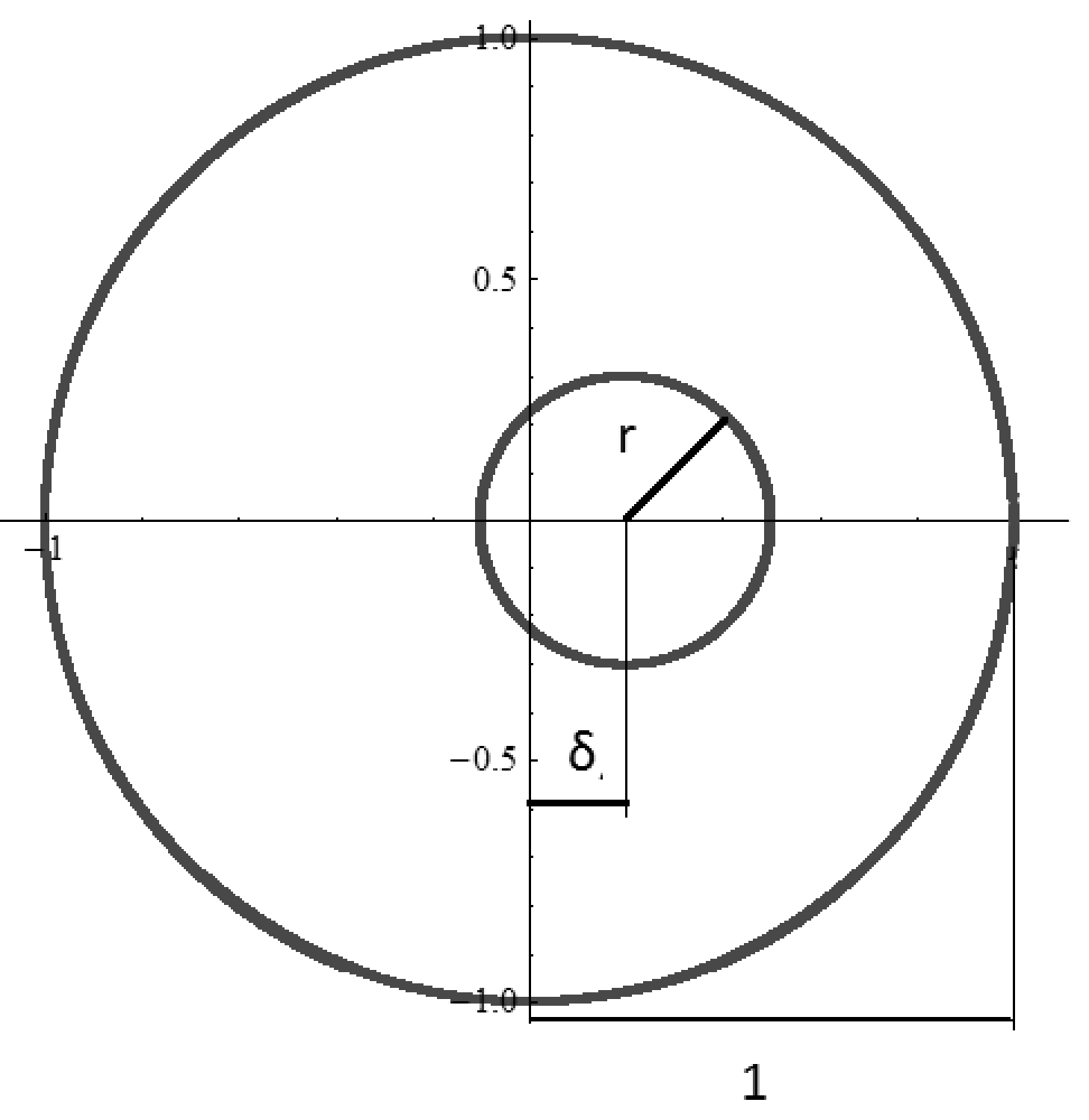}
\end{center}
\caption{\label{f1} Boundary shape of the annular billiard. The parameter $\delta$ is considered to be positive.}
\end{figure}

Among the systems demonstrating chaoticity, one of the simplest family is presented by plain billiards, which are two-dimensional single-particle potential-free systems, limited in finite domain. They are defined by two-dimensional single-particle Hamiltonian
\[H=\left\{
\begin{array}{cl}
\frac{p_x^2+p_y^2}{2m}, & \mathrm{for}\ {{\bf r}\in\Omega};\\
 & \\
\infty, & \mathrm{for}\ {{\bf r}\notin\Omega}.
\end{array}
\right.\]
The dynamics (both the classical and quantum) in such systems is defined by solely the geometry of the boundary. In the present paper the boundary is chosen to have the following form: motion takes place in the region limited by two circles of different radii (see Fig.\ref{f1}). Such system is called the annular billiard.
In the case $\delta=0$ (centers of the bounding circles coincide) the system has two independent integrals of motion: energy and angular momentum. As motion takes place in two-dimensional space, it is sufficient to make the dynamics purely regular. For $\delta>0$ (inner disk is shifted relative to the external one) the system looses rotational symmetry and the angular momentum is not already conserved, thus the system possesses mixed (chaotic and regular) dynamics. The annular billiard is a well-known system widely used in the present context \cite{bohigas,doron,robinett}. Besides the convenience for analysis of fundamental properties of quantum chaos, the annular billiard represents interest for experimental and applied researches, in frames of which it models the microwave cavities \cite{richter,backer}.

A characteristic feature of chaotic system which manifests in quantum dynamics is the dynamical tunneling. This notion generalizes usual tunneling. The dynamical tunneling represents a transition between two quantum states which is forbidden for their classical counterparts for some reason (not obligatory because of static potential barrier) \cite{kesha}.

In 1993 year Bohigas and colleagues \cite{bohigas} were first to describe the dynamical tunneling in the annular billiard. There is a family of trajectories in this system which do not touch the inner disk. They are usually called the whispering gallery modes. As those trajectories do not feel the inner disk, they are regular and quasiperiodic. In his paper Bohigas demonstrates the transition between the two states, corresponding to two whispering gallery orbits. This phenomenon has a distinct peculiarity: the tunneling rate varies in many orders of magnitude with comparably small changes of system parameters.

\begin{figure}
\includegraphics[width=\columnwidth]{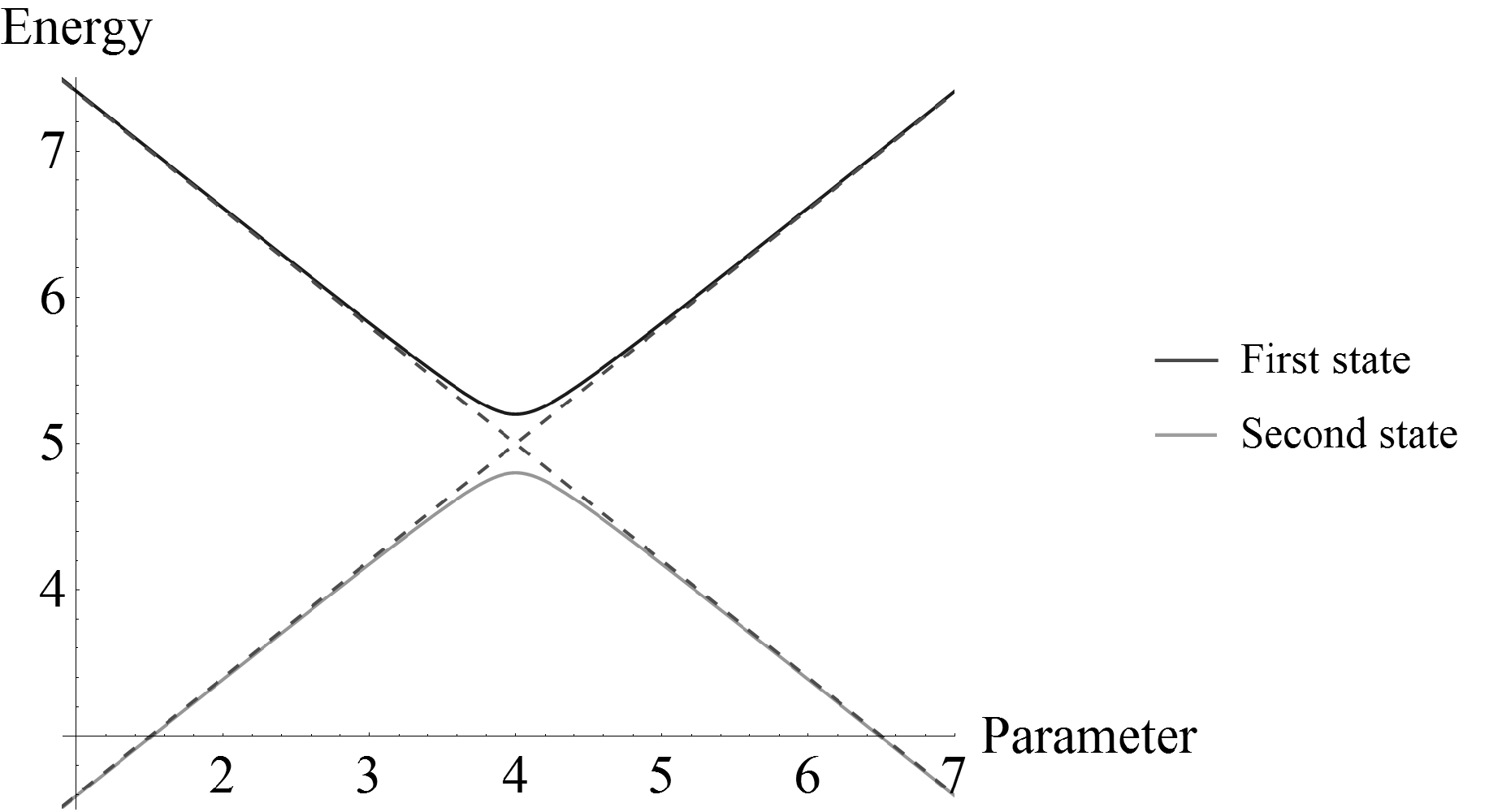}
\caption{\label{f2} Example of the level repulsion. A typical dependence on a parameter of Hamiltonian is shown for two energy levels of the states of identical symmetry type.}
\end{figure}

One more important feature of chaotic quantum systems is the presence of level quasicrossings (level repulsions, avoided crossings). The theorem proven by Wigner and von Neumann \cite{wigner} states that two initially different values of observables, which correspond to states in the same symmetry type, can be made equal by variation of Hamiltonian parameters at most on a manifold of dimension $N-2$, where $N$ is number of the parameters varied. Therefore variation of only single parameter cannot lead to intersection of energy values of two levels and thus it forms a hyperbola in vicinity of the nearing point (see Fig.\ref{f2}). The quasicrossings are related to such properties of quantum chaos as changes in the level statistics and chaos-assisted tunneling \cite{doron}. It was one of the manifestations of the latter effect which was described in the Bohigas' work.
There remains an open question about the details of quantum-classical correspondence in the case of the mixed type of dynamics. It follows from qualitative considerations that the wave functions of the states in such systems corresponds to some invariant sets of classical dynamics. However such assumptions does not hold near the quasicrossing points on the intermediate stage of the mode exchange. In the present work we make a detailed analysis of localization features for (quasi-)distributions in the phase space of quantum states on the classical trajectories of the system. For that purpose we use the concept of the Husimi-Poincar\'e functions. We consider a quasicrossing between a regular and a chaotic state. Besides that we explicitly demonstrate tunneling between them near the quasicrossing point.
\section{Problem Setup and Numerical Method Description}
A quantum analog of classical billiard represents a particle freely moving in a region $\Omega$ bounded by elastic walls. It is formally expressed by the following stationary Schr\'odinger equation
\begin{equation}\label{sse}
E\psi({\bf r})=-\frac{\hbar^2}{2m}\Delta\psi({\bf r}),\quad {\bf r}\in\Omega,
\end{equation}
with the Dirichlet boundary condition
\begin{equation}\label{dbc}
\psi({\bf r}\in\partial\Omega)=0,
\end{equation}
where $\partial\Omega$ denotes boundary of the region $\Omega$.

Non-trivial shapes of boundary of the domain $\Omega$ do not allow analytical solutions of the equation (\ref{sse}). Therefore it is necessary to apply numerical methods. First for convenience let us rewrite the equation (\ref{sse}), substituting the energy by the wave number squared (below and further we use the system of units with $2m=\hbar^2=1$):
\[(k^2+\Delta)\psi({\bf r})=0,\quad {\bf r}\in\Omega.\]
There are plenty of numerical methods for solution of Schr\"odinger equation, however not all of them can be efficiently applied to quantum billiards. The main feature of such systems---absence of the potential---can be used for elaboration of efficient algorithms for the numerical solution. A widely used family of such algorithms is presented by the boundary integral method. Its efficiency is based on utilization of the fact that the eigenfunction of the equation (\ref{sse}) is completely determined by the values of normal derivative on the boundary of the billiard. Let us introduce the normal derivative as a function of the natural length parameter $s$ along the boundary:
\[u(s)\equiv\frac{\partial}{\partial n}\psi({\bf r}(s)),\quad {\bf r}(s)\in\partial\Omega.\]
The wave function in arbitrary point is expressed through the normal derivative on the boundary in terms of the Green function of the equation (\ref{sse}) in the following way \cite{backer2}:
\[\psi({\bf r})=\oint\limits_{\partial\Omega} G_k({\bf r},{\bf r}(s))u(s)ds.\]
Following the paper \cite{robnik} we may transform Schro\''odinger equation for our quantum billiard into an integral equation by means of the {\it regularized} Green function $G_k({\bf r},{\bf r}')$
which solves the following defining equation:
\begin{equation}\label{EquationGreenFunctionRegular}
(\Delta+ k^2 )G({\bf r},{\bf r'})=\delta({\bf r}-{\bf r'})-\delta({\bf r}-{\bf r'_R}),
\end{equation}
where ${\bf r}$ and  ${\bf r'}$ are in $ {\cal B}\cup {\partial {\cal B}}$,
and $\bf r'_R$ is the mirror image of ${\bf r'}$ with respect to the
tangent at the closest lying point on the boundary, and thus if ${\bf r'}$
is sufficiently close to the boundary then ${\bf r'_R}$ is outside the
billiard ${\cal B}$.
The solution can easily be found in terms of the free propagator (the free
particle Green function on the full Euclidean plane)
\begin{equation}
G_0({\bf r}, {\bf r'}) = - \frac{1}{4} i H_{0}^{(1)}(k|{\bf r}-{\bf r'}|),
\label{eq:Gfree}
\end{equation}
where $H_{0}^{(1)}$ is the zero order Hankel function of the first kind
\cite{as}, namely
\begin{equation}
G({\bf r},{\bf r'}) = G_{0}({\bf r},{\bf r'}) -G_{0}({\bf r},{\bf r'_R}),
\label{eq:Green}
\end{equation}
such that now $G({\bf r},{\bf r'})$ is zero by construction for any
${\bf r'}$ on the boundary.

With the use of the normal derivative the equation (\ref{sse}) with the boundary condition (\ref{dbc}) can be rewritten in form of the integral equation
\begin{equation}\label{u_s}
u(s)=-2\oint\limits_{\partial\Omega} \frac{\partial}{\partial n} G_k({\bf r}(s),{\bf r}(s'))u(s')ds'.
\end{equation}
For the normal derivative of the Green's function one obtains
\[\frac{\partial}{\partial n} G_k=\frac{ik}{4}\frac{{\bf n}(s)\cdot({\bf r}(s)-{\bf r}(s'))}{|{\bf r}(s)-{\bf r}(s')|}\times H_1^{(1)}(k|{\bf r}(s)-{\bf r}(s')|).\]
Therefore the two-dimensional problem is reduced to the one-dimensional one which is much easier to solve numerically. For that purpose we present the boundary integral in the equation (\ref{u_s}) in form of finite sum. We thus obtain a finite system of uniform linear equations
\begin{equation}\label{ule}
\sum\limits_k A_{ik}u_k=0,
\end{equation}
where where $A_{ik}$ is a dense, complex non-Hermitean matrix \cite{backer2}. Solution of the equation (\ref{ule}) requires to find those values of the wave number $k$ which turn the system determinant $\det(A)$ to zero. It is the values of $k$ that give the required stationary energy levels.

A difficulty arises here: there are no real values of $k$ which make the obtained determinant equal to zero exactly, because it is a complex function of a complex argument. Due to the discretization of the integral (\ref{u_s}) the determinant $\det(A)$ will not become zero but only close to zero \footnote{actually, the discretization shifts the zeros slightly away from the real axis, see \cite{backer58,backer59}}. Besides that, calculation of the wave functions requires to find the normal derivative $u(s)$ in the discretizing points of the boundary. To get rid of this problems one should in advance to make the singular decomposition of the matrix $A$
\[A=USV^\dag,\]
where $U$ and $V$ are unitary matrices, and $S$ is a diagonal real matrix composed of singular values of matrix $A$. Those singular values are defined as eigenvalues of the self-adjoint operator $AA*$. Minima of the first (minimal) singular value as function of the wave number correspond to the sought eigenvalues, and the corresponding columns of the matrix $U$ give values of the normal derivative. Besides that, such approach allows to detect the energy values which lie near the quasidegenerate states. If both the first and second singular values have sharp minima near the obtained eigenvalue then there are in fact two states in close vicinity of the energy eigenvalue.

\begin{figure}
\includegraphics[width=\columnwidth]{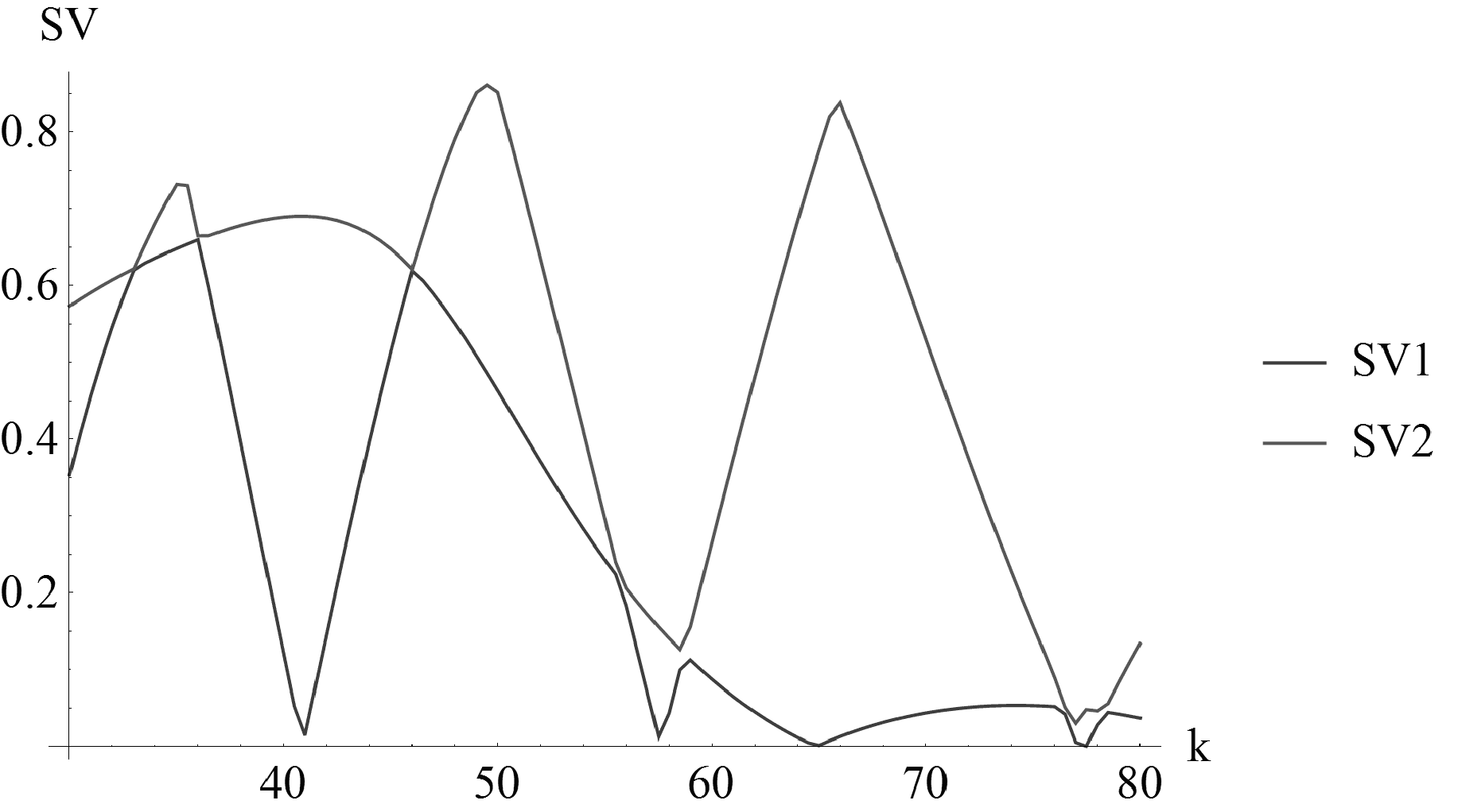}
\caption{\label{f3} Dependencies of first and second singular values on the wave number in symmetric annular billiard ($r=0.2$). One can see that near the value $k=77$ both the singular values have a sharp minimum, which gives evidence for presence of a multiplet.}
\end{figure}

\begin{figure}
\includegraphics[width=\columnwidth]{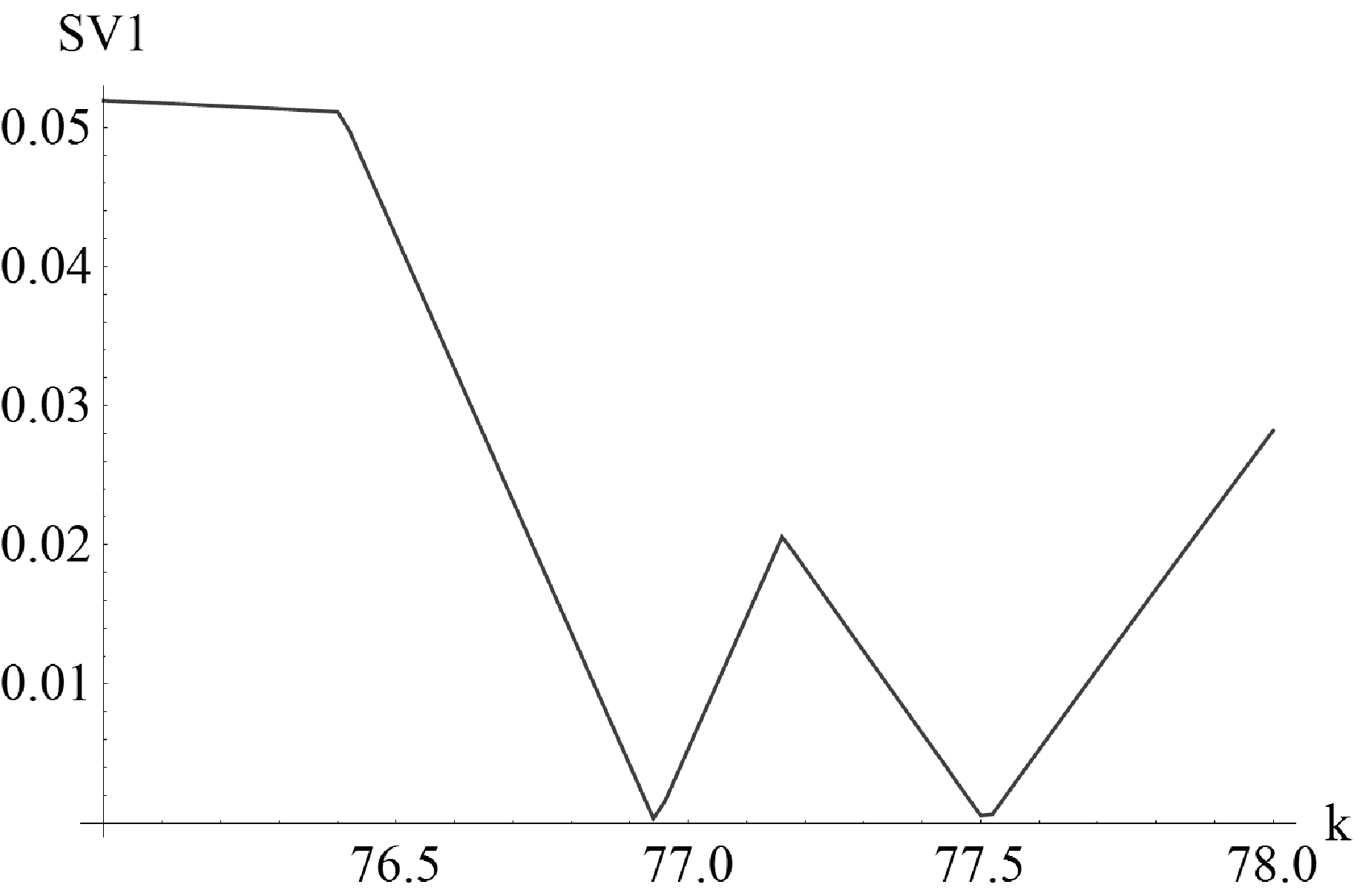}
\caption{\label{f4} The first singular value calculated with improved accuracy is shown here in the vicinity of minimum of the second singular value (see Fig.3). It can be seen that one can reveal two values of the wave number $k$ by decreasing of the calculation step and increasing of the accuracy.}
\end{figure}

\begin{figure}
\includegraphics[width=\columnwidth]{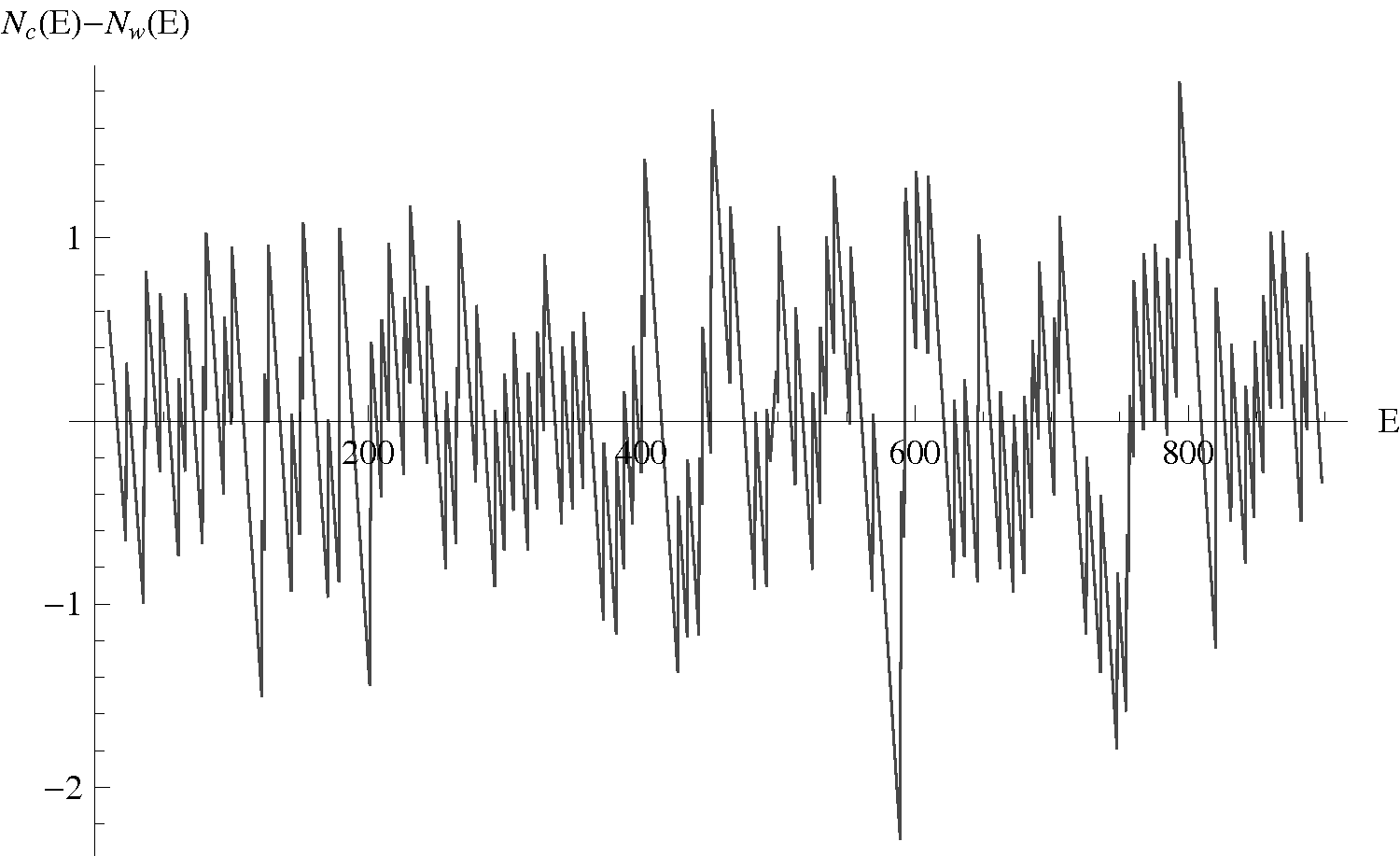}
\includegraphics[width=\columnwidth]{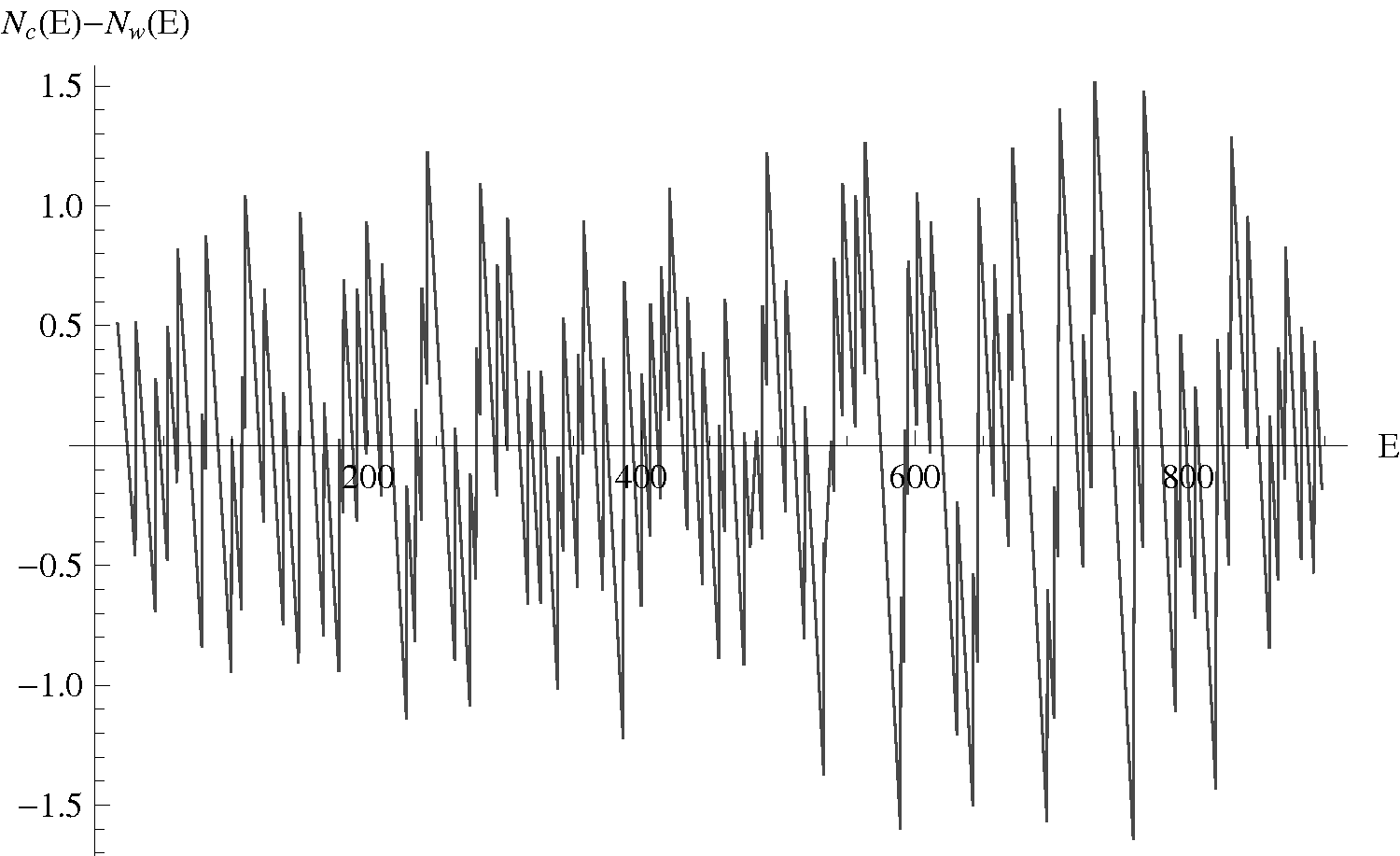}
\caption{\label{f5} The fluctuating part of the cumulative distribution function for symmetric (upper) and antisymmetric (lower) states obtained numerically.}
\end{figure}

Examples of calculation of the singular values are presented on Figures \ref{f3} and \ref{f4}. One can easily see from this example that usage of the singular decomposition allows to detect the quasicrossings and find all the quasidegenerate levels  with considerable saving of the computational efforts. First of all one should find all minimum values of the first singular value. In the case of detection of minimum for the second singular value one repeatedly performs search for required values of the wave number with ever decreasing step and higher discretization degree of the matrix $A$, aimed to improve the accuracy.
To verify the obtained results we use the level statistics. Let us introduce the integral level density---the state number staircase function, which counts the number of levels with energy not greater than the given one
\begin{equation}\label{n_e}
N(E)=\{n|E_{i\le n}\le E, E_{n+1}>E\}.
\end{equation}
For billiards one can use the generalized Weyl formula which gives the averaged smooth component of the function (\ref{n_e}) \cite{balt}:
\begin{equation}\label{n_w}
N_w(E)= \frac{S}{4\pi}E-\frac{L}{4\pi}\sqrt{E}+C.
\end{equation}
Here we dropped the terms of higher orders of smallness, $S$ is the interior area of the domain $\Omega$, $L=L_D-L_N$, $L_{D/N}$ are the parts of the boundary with the Dirichlet and Neumann conditions respectively. The desymmetrization we use here allows to consider separately the level statistics for states of different symmetry. At all that one should take into account that the part of the boundary along the symmetry line of the billiard has Neumann boundary conditions for symmetric states and Dirichlet boundary conditions for the antisymmetric ones. Subtracting the smoothed part (\ref{n_w}) from the numerically obtained integral level density $N_c(E)$, one obtains the fluctuating part. In the case when a level is missing in the numerical results it can be immediately detected by strong deviation from zero of the average value of the fluctuation component. Fig.\ref{f5} shows the obtained dependencies, which give evidence for completeness of the calculated spectrum.
\section{Quasicrossings of Quantum States and Dynamical Tunneling}
Consider an initially regular annular billiard with $\delta=0$. Let us start gradually shift the inner disk, i.e. to increase $\delta$. The tori corresponding to orbits near the inner disk will be destroyed and distorted. The quantum states corresponding to them will get all their properties considerably changed, in particular their energy significantly changes with variation of $\delta$. However as Lazutkin have shown \cite{berry}, in the case of smooth billiard boundary\footnote{more precisely, the boundary must have continuous derivatives with respect to the parameter up to the $553$-th one} there are non-destroyed tori remaining in the system for any set of parameters. In the considered case of the annular billiard there are whispering gallery orbits up to $\delta=1-r$. As the whispering gallery tori in classical dynamics absolutely do not feel the variation of the $\delta$ (up to certain its value), it is then natural to assume that the semiclassical quantum states corresponding to the tori will be extremely weakly sensitive to the variations of $\delta$ too.

Values of some energy levels get closer at variation of $\delta$. If these levels have the same symmetry then after certain nearing they start to repel and form a quasicrossing. While passing the quasicrossing region, these states exchange the so-called adiabatic properties. We are interested in peculiarities of this exchange in the case of quasicrossing between chaotic and regular levels. The simplest task is to detect such mode exchange between the states which correspond to classical orbits of whispering gallery and chaotic ones. This is due to the fact that, as was mentioned before, the energy levels of the former weakly depend on the shift parameter, while the letter depend much stronger.

\begin{figure}
\includegraphics[width=\columnwidth]{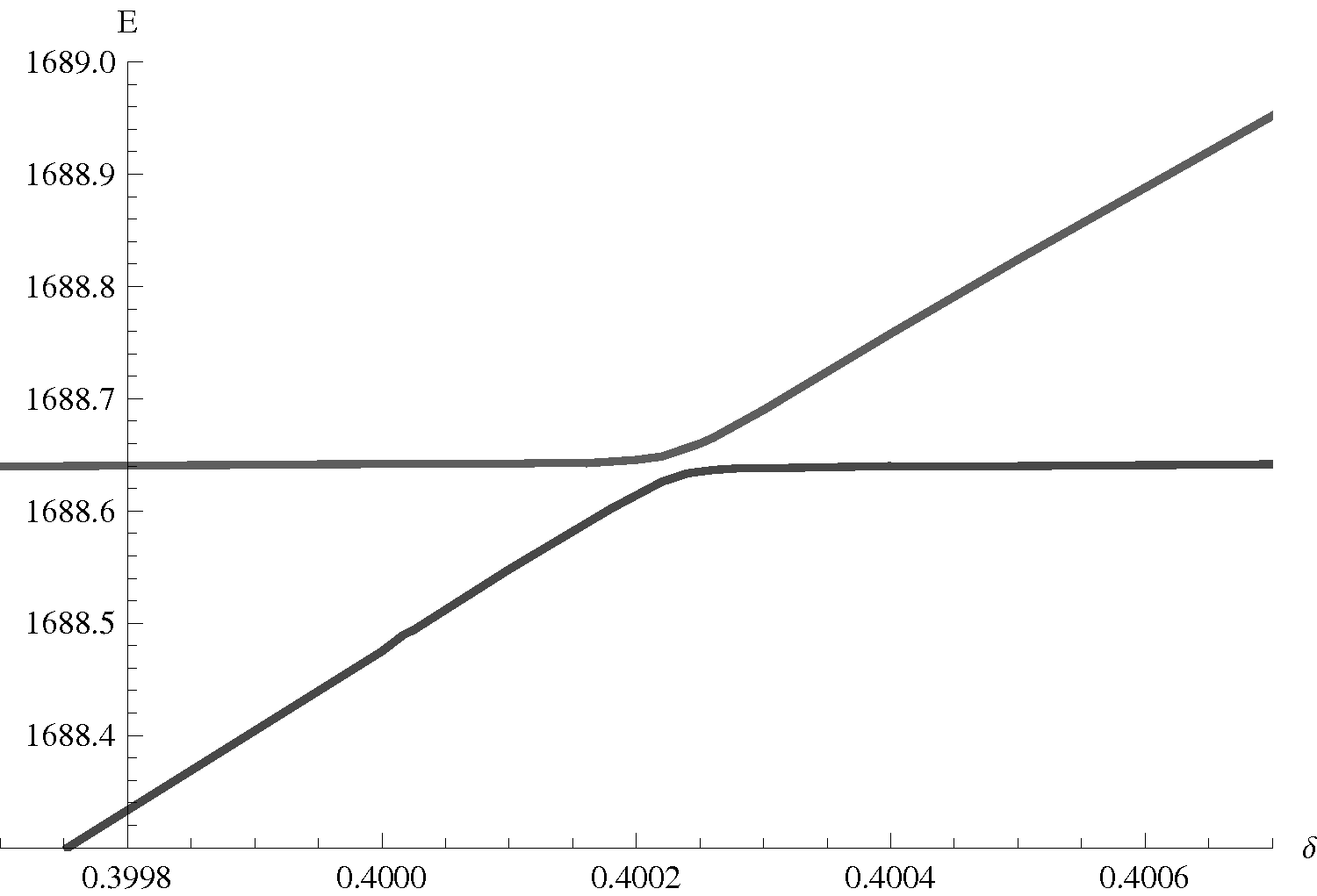}
\caption{\label{f6} Dependence on the shift parameter $\delta$ for two energy levels of antisymmetric states. It is an example of level repulsion in the annular billiard.}
\end{figure}

Figure \ref{f6} presents the dependence of energies on parameter $\delta$ for two antisymmetric states. The very mode exchange takes place in very narrow range of energy and parameter space (of order $10^{-4}$). Thus numerical results with lower accuracy may look as real level crossing. Near the quasicrossing the energy level lines have characteristic hyperbolic shape.
As can be seen, one of the levels has inherently practically constant energy value, the other considerably shifts with growth of $\delta$. One can thus assume that the former state corresponds to the whispering gallery and the latter---to the state related to chaotic sea of classical dynamics.

\begin{figure}
\includegraphics[width=\columnwidth]{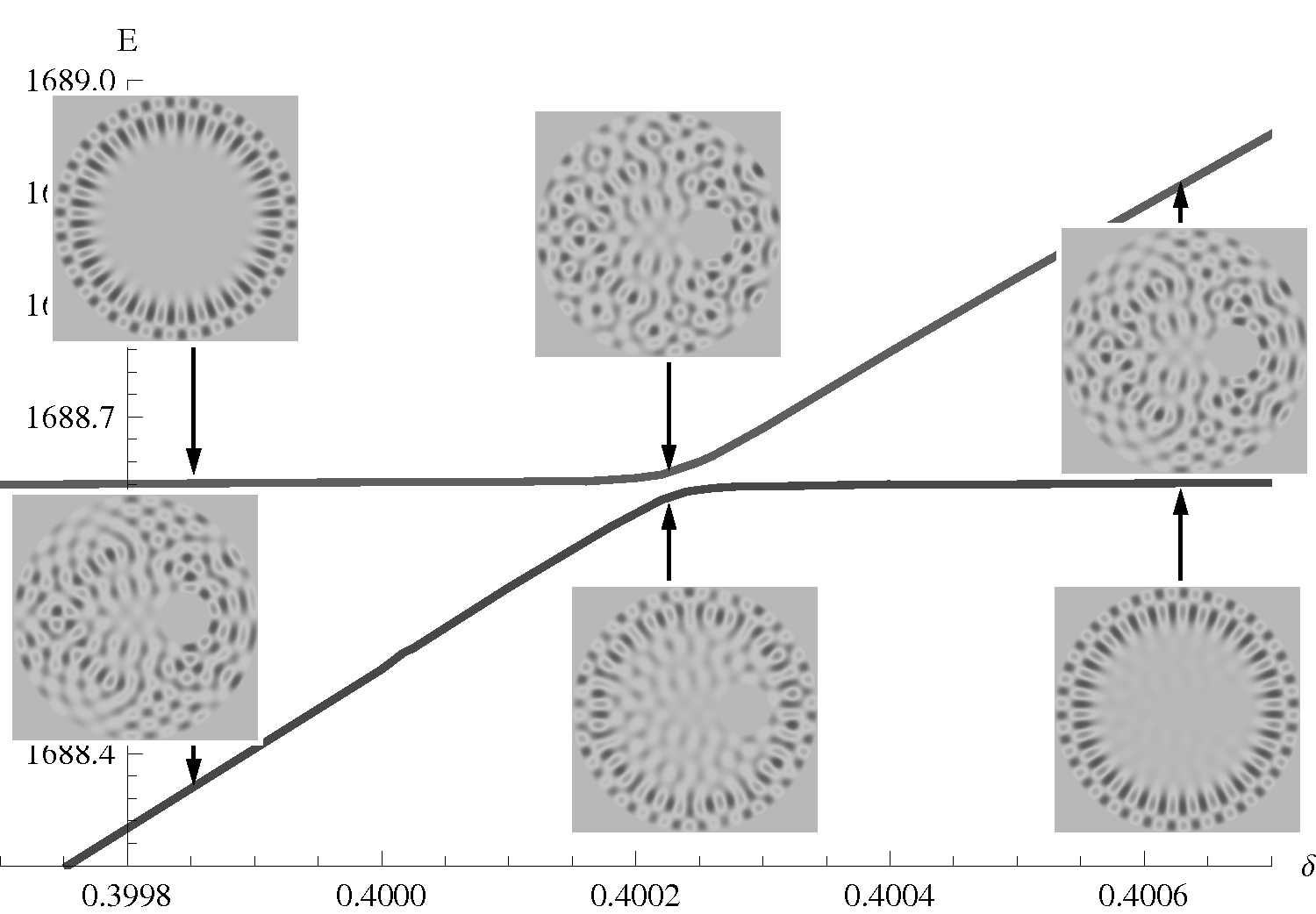}
\caption{\label{f7} The same as on Fig.6 with addition of images of wave functions for some states (the corresponding energy values are shown by arrows).}
\end{figure}

Figure \ref{f7} shows wave functions of the above described states. Let us first consider the wave functions of the states in relative distance from the quasicrossing. First of all it should be noted that the modes under investigation completely exchange the wave functions. Distinction of the wave functions to the left and to the right of the quasicrossing is extremely weak and it is related to variation of the parameter $\delta$. The state corresponding to the horizontal line has regular shape of the wave function as expected. This wave function is localized near the outer disk, it strongly decreases near the inner one, which shows its connection to the whispering gallery orbits. The other wave function, also as expected, is clearly chaotic, what is supported for example by irregularity of its nodal lines. More rigorous correspondence of wave functions of these states to the classical structures will get more detailed investigation in the next section.
Wave functions near the quasicrossing point were calculated for parameter value which provides the minimum of the energy splitting. It can be seen that the states are close to the original ones but they are mixed in some sense. The lower medium state on the Figure \ref{f7} is clearly close to the regular one, which is far from the quasicrossing point. Nevertheless its wave function is substantially distorted. There are traces of the chaotic wave function clearly visible in the central part.

It is natural to assume that during the mode exchange the wave functions represent linear combinations of the original ones. However this notion cannot be understood too literally here, because variation of the parameter $\delta$ affects the configuration space itself. But smallness of the parameter changes allows to verify such a hypothesis. Let us construct a linear combination of the wave functions near the quasicrossing point. The coefficients should be fitted in order to make the obtained combination similar to one of the original wave function. For example, it can be seen from the Fig.\ref{f7} that the wave function corresponding to the whispering gallery mode has a nodal line along the vertical diagonal of the outer boundary. Minimization of the absolute value of the considered combination on this line gives a combination which is practically identical to the original wave function\footnote{taking into account the correction due to the shift of $\delta$}. The state orthogonal to the latter in turn recovers the original chaotic one.

\begin{figure}
\includegraphics[width=0.45\columnwidth]{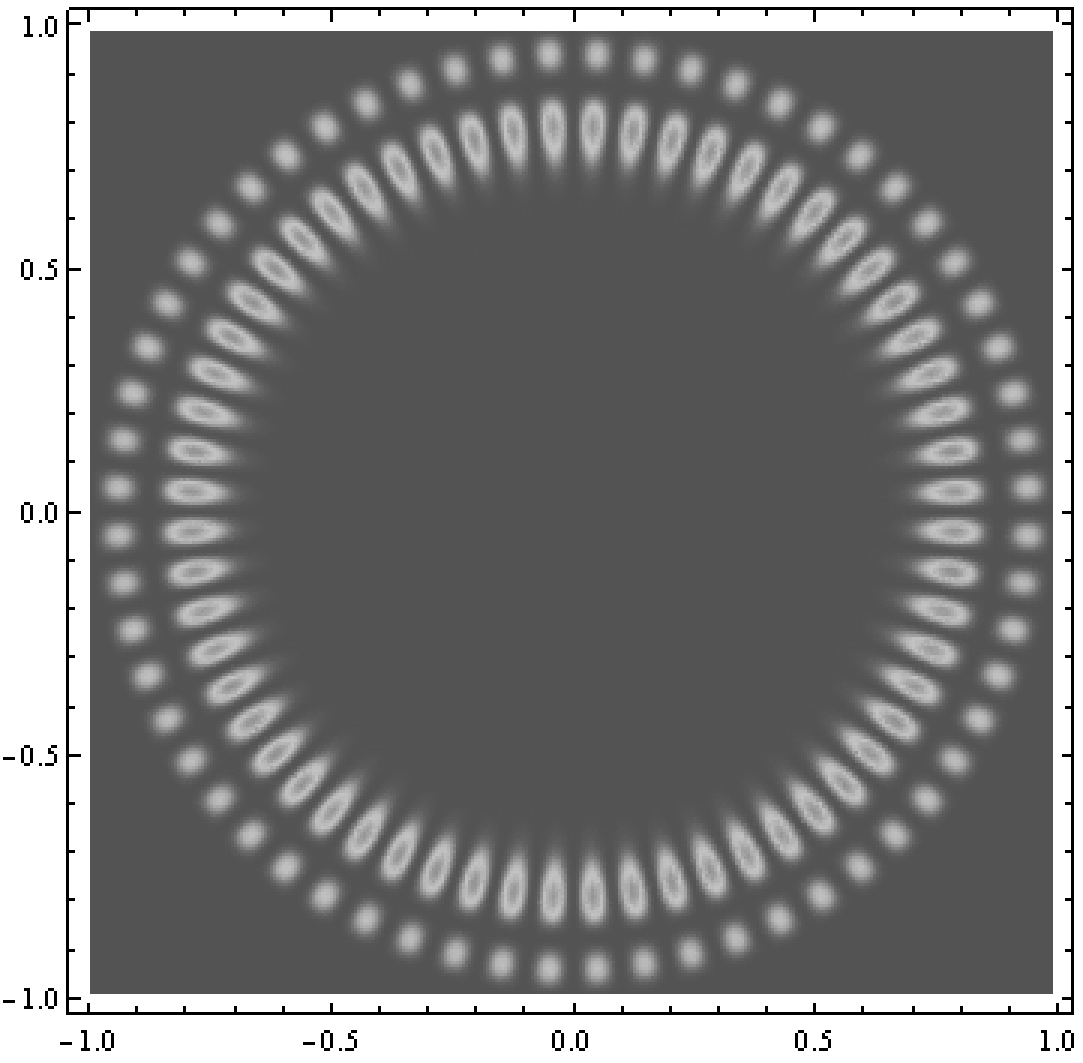}
\includegraphics[width=0.45\columnwidth]{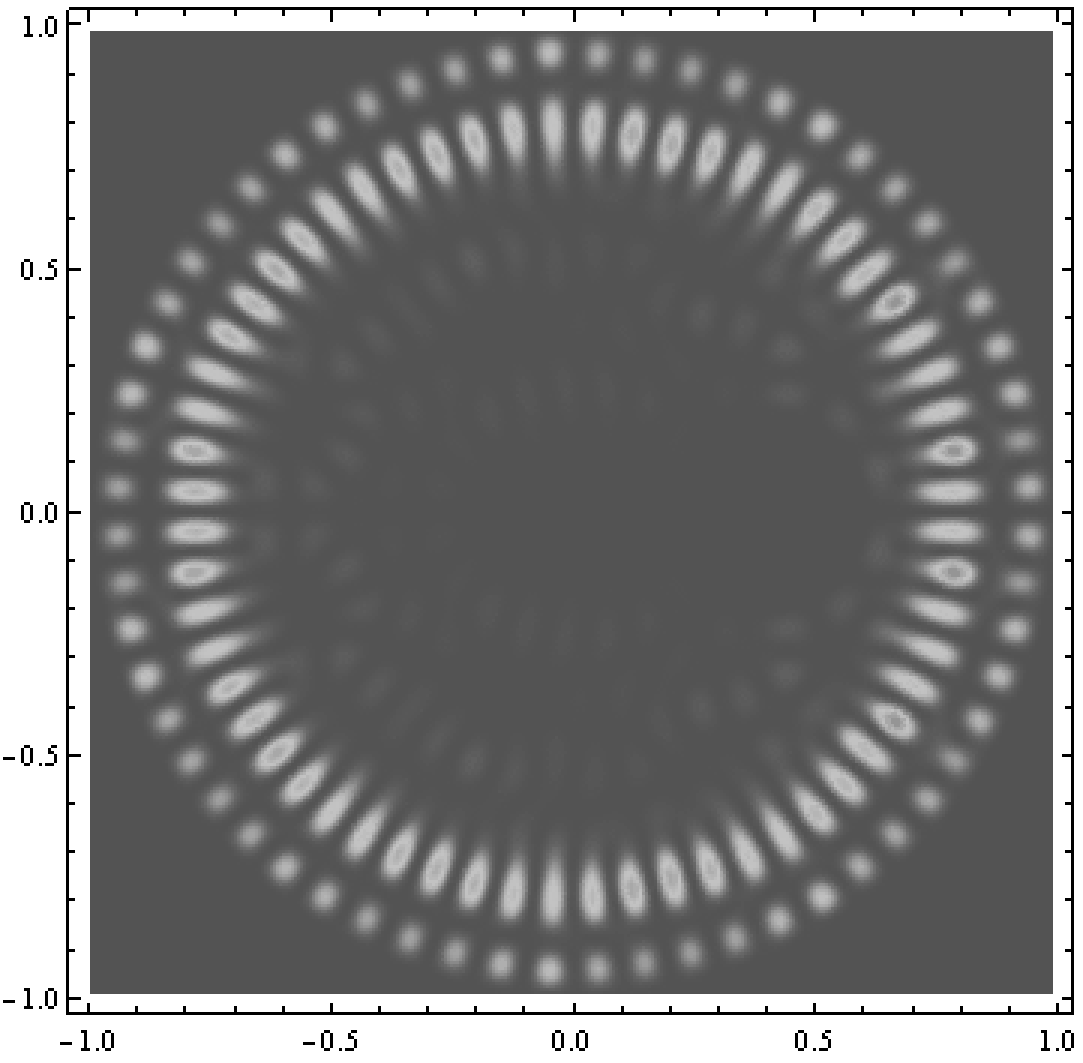}\\
\includegraphics[width=0.45\columnwidth]{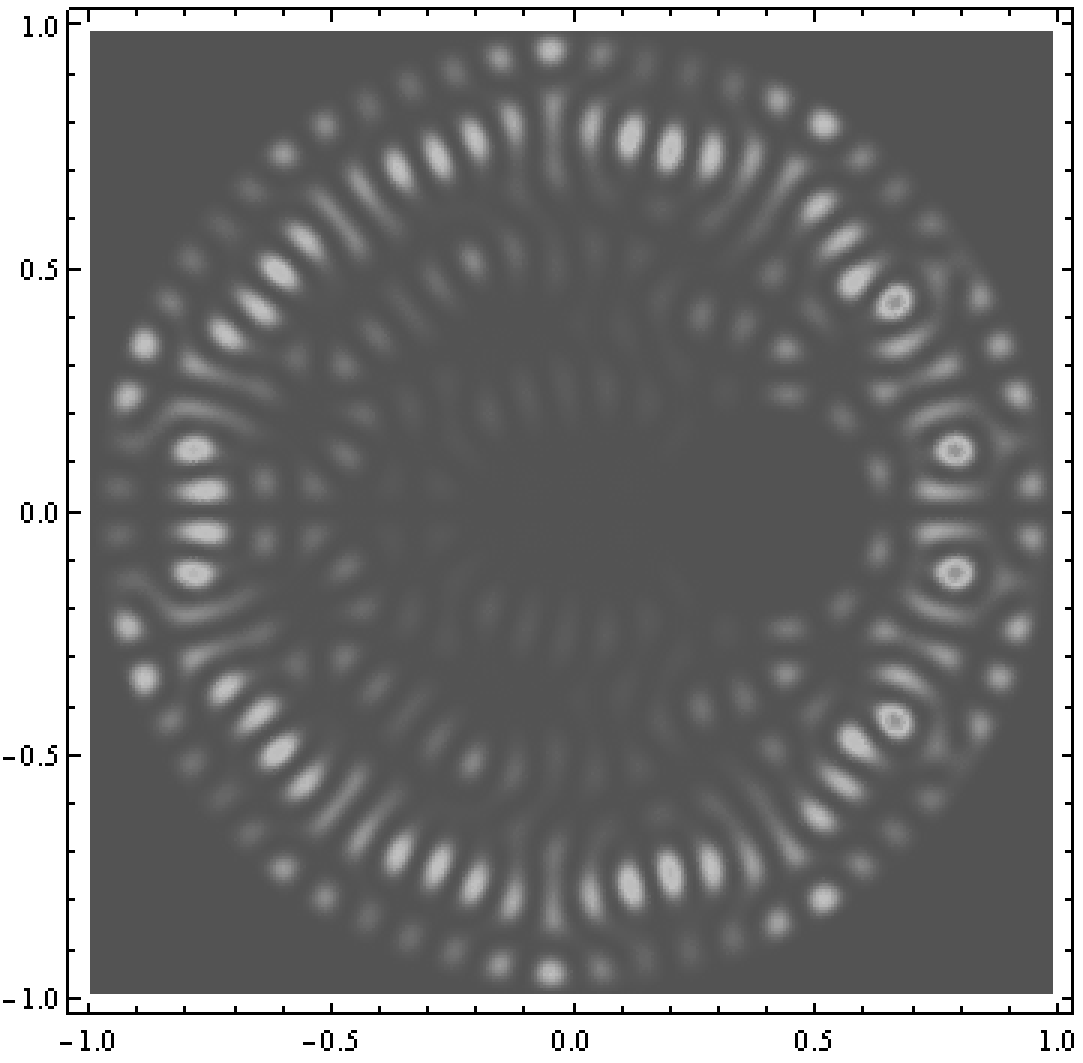}
\includegraphics[width=0.45\columnwidth]{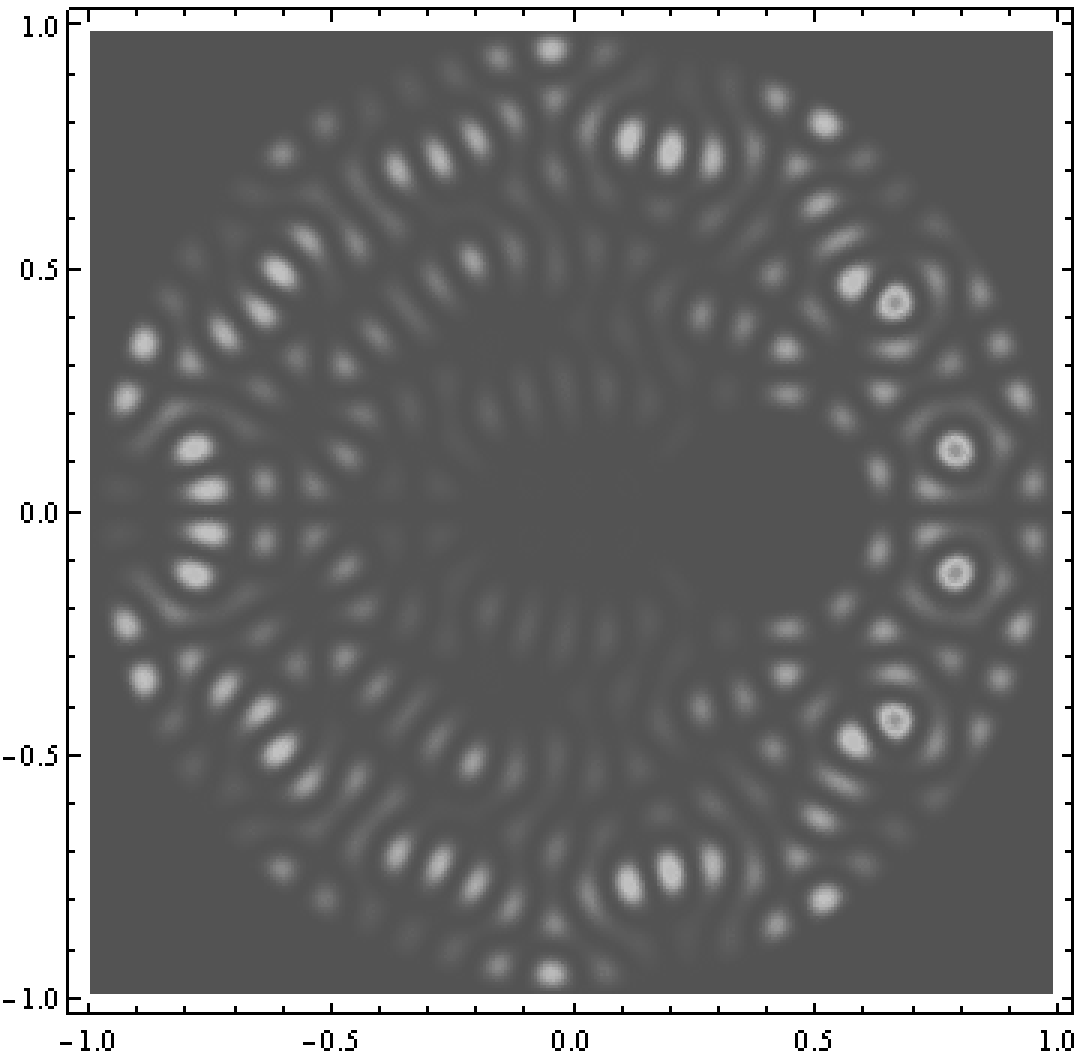}
\caption{\label{f8} Consecutive stages of wave function evolution.}
\end{figure}

Having obtained the chaotic and regular combinations of the wave functions, one can observe the dynamical tunneling between these states. In the billiard with the parameter $\delta$ close to the quasicrossing point we prepared the regular wave packet described above. Figure \ref{f8} shows the time evolution of this ave packet. As can be seen, the regular state (partially) tunnels into the chaotic one. It is initially manifested as the wave function, initially concentrated on the boundary, (partially0 spreads over whole configuration space. Distraction of the initial rotational symmetry takes place. The nodal lines loose regularity, forming new structures and distorting the remaining ones. After that the states returns to the original one.
\section{Peculiarities of Quantum-Classical Correspondence}
We need to relate the quantum states to the classical dynamics. Different quasi-distributions of quantum states in phase space represent a convenient tool for this purpose. It is impossible to construct actual distribution because of a number of restrictions imposed by the uncertainty principle. Therefore one have to use different variations in some sense similar to the distribution function. In the present work we shall use the concept of the Husimi functions. It satisfies all the criteria required for a distribution functions except the axiom of additivity of probabilities in terms of Kolmogorov axiomatic.

The Husimi function is defined (up to a coefficient independent on phase variables) as the absolute value squared of the folding of the wave function with the coherent state\footnote{the state of minimum uncertainty}. However it is problematic to work with complete Husimi functions because the visualization in four-dimensional space presents a difficulty.

To solve this problem let us address the classical dynamics. Absolute value of momentum of a particle, as well as its energy,  does not affect shape of trajectory of its motion in a billiard. Only direction of the motion and position of the particle matters. Moreover motion of the particle is completely determined by position of its last collision with the boundary and value of the incident angle. Therefore motion of the particle can be reduced to discrete mapping in coordinate space defined by the pair (position of the collision with the boundary, cosine of the incident angle). This pair is called the Birkghoff coordinates. Such mapping represents in fact a kind of Poincar\'e map.

As was mention before, the quantum dynamics in billiards can be completely reduced to that of normal derivative of the wave function on the boundary. This fact shows an analogy with classical dynamics. Let us construct an analogue for Husimi function for the normal derivative on the boundary---the so-called Husimi-Poincar\'e function. For that purpose one has to make further improvements. First of all it is necessary to provide periodicity of the function on the boundary, i.e. take into account its closure. For that purpose we introduce an one-dimensional coherent state which is periodic on the outer boundary \cite{backer3}
\begin{equation}\label{coh}
c_k(q,p;s)\equiv\left(\frac k\pi\right)^{1/4}\sum\limits_{m=-\infty}^\infty e^{ik\left[p(s-q+2\pi m)+\frac i2(s-q+2\pi m)^2\right]},
\end{equation}
where $k$ is the wave number of the state, which the Husimi-Poincar\'e functions is calculated for; $s$ is the natural length parameter along the boundary; $(q,p)\in[0,2\pi)\times R$ are arguments of the function.

Calculation of folding of the two functions on the boundary is quite arbitrary as the metrics of the two-dimensional domain does not induce any metrics on the boundary. Let us define the scalar product on the boundary in the simplest way: as an integral with unit kernel. Taking projection of the normal derivative of the wave function for a given state $u(s)$ with the coherent state (\ref{coh}) one obtains the Husimi-Poincar\'e function
\[h(q,p)=\frac1{2\pi k}\left|\int\limits_0^{2\pi}c_k^*(q,p;s)u(s)ds\right|^2.\]

The coordinates $(q,p)$ represent direct analogy t the Birkghoff coordinates. Having construct the classical Poincar\'e map and compared it with the calculated Husimi-Poincar\'e functions, one can determine which classical object correspond to certain quantum state.

\begin{figure}
\includegraphics[width=\columnwidth]{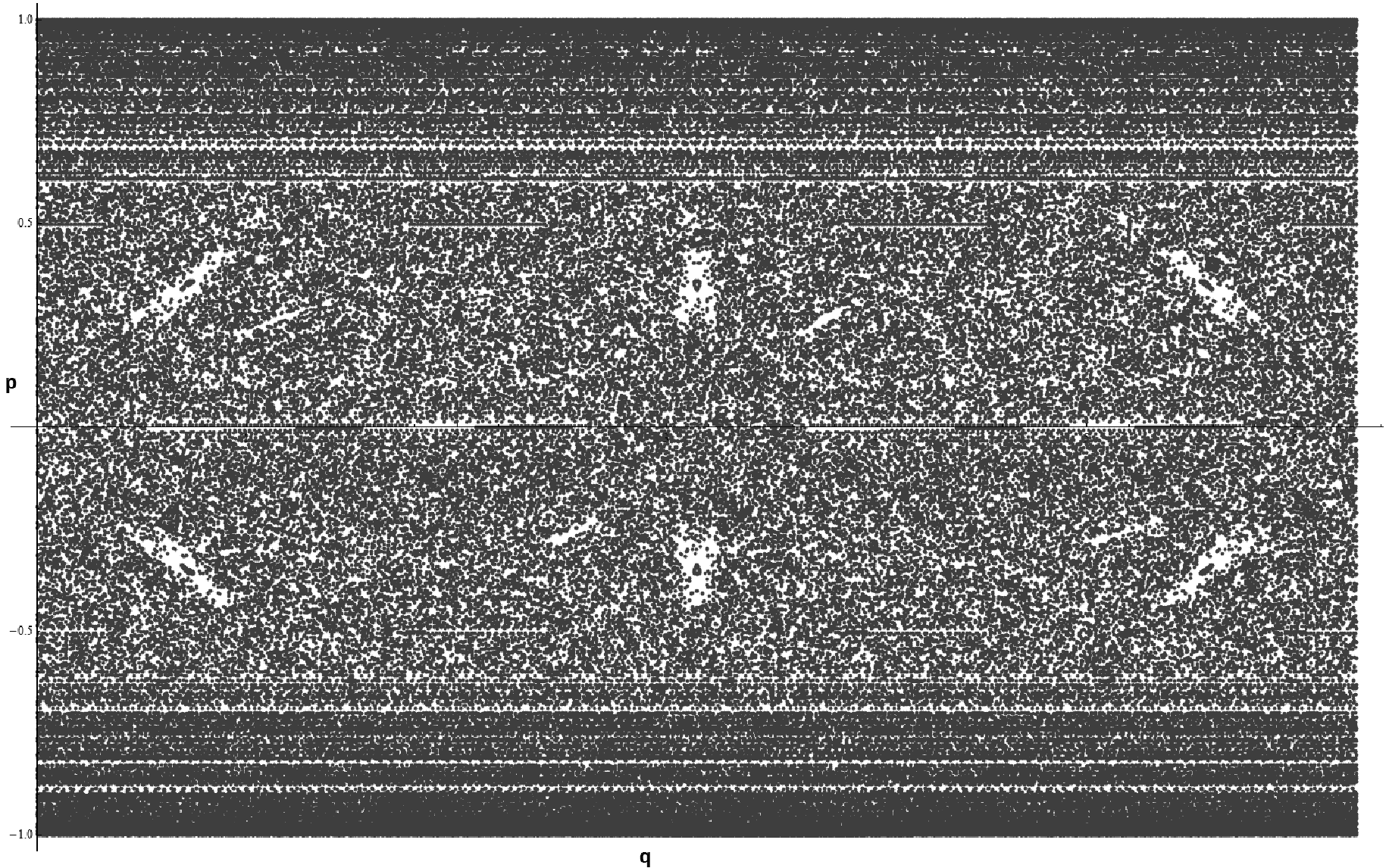}
\caption{\label{f9} Poincar\'e surface of section (the stroboscopic map) for the annular billiard.}
\end{figure}

\begin{figure}
\includegraphics[width=0.45\columnwidth]{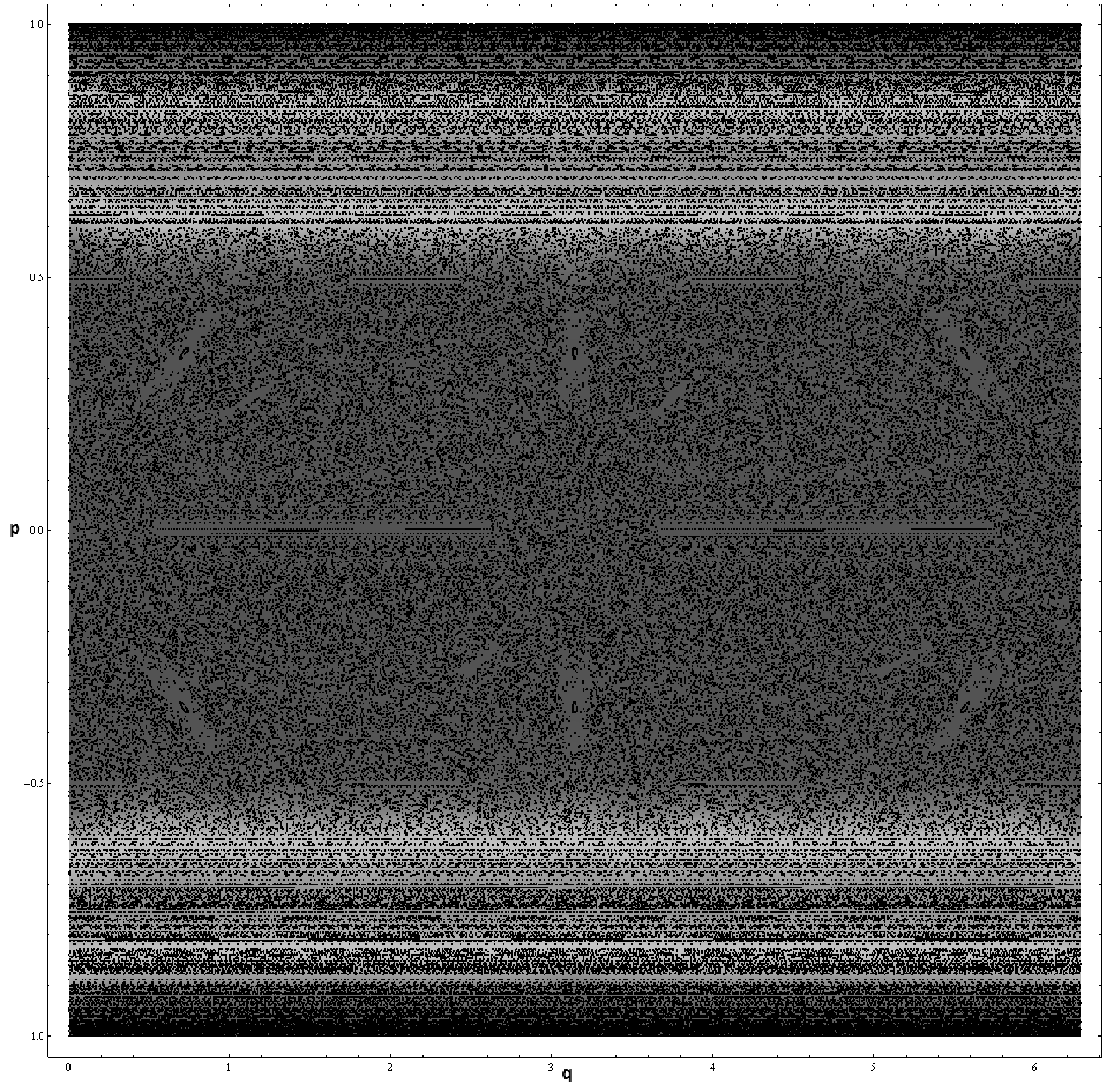}
\includegraphics[width=0.45\columnwidth]{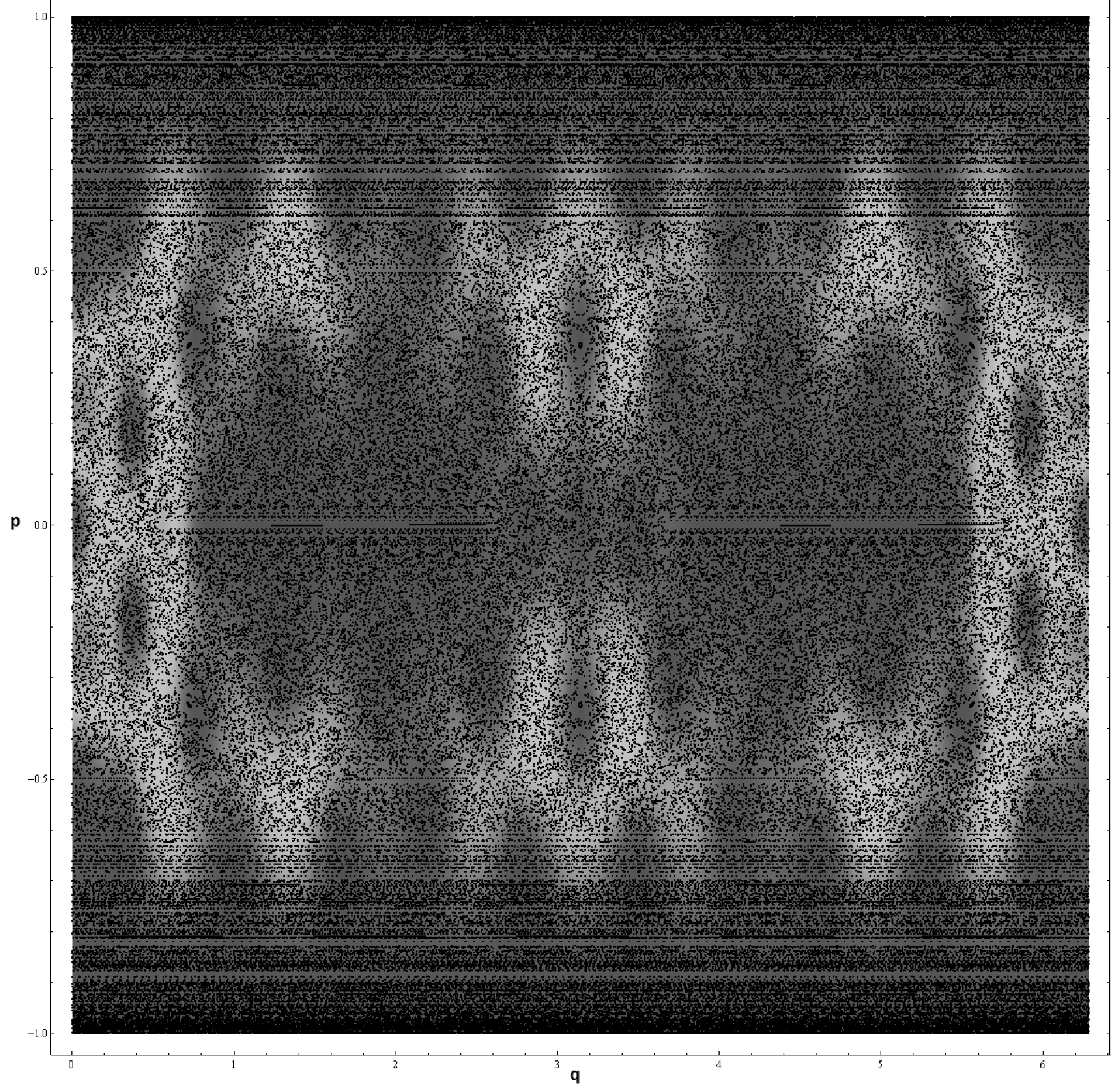}
\caption{\label{f10} The Husimi-Poincar\'e distribution functions for two states (the left ones on Fig.\ref{f7}) mapped over the classical Poincar\'e surface of section.}
\end{figure}

\begin{figure}
\includegraphics[width=\columnwidth]{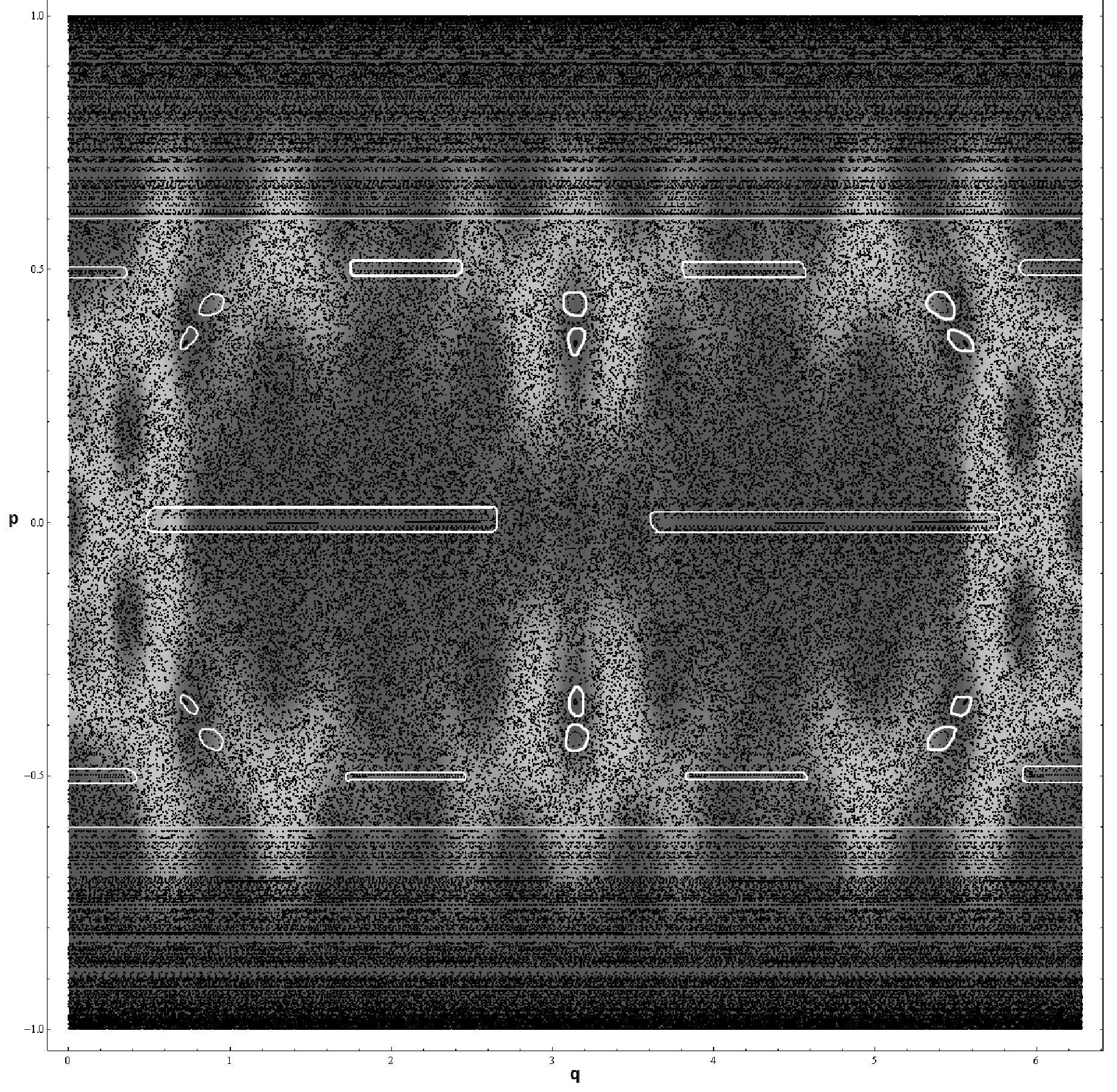}
\caption{\label{f11} The Husimi-Poincar\'e distribution function for a state near the level repulsion point (the upper middle state on Fig.\ref{f7}) mapped over the classical Poincar\'e surface of section. Separate solid lines mark the boundaries of the whispering gallery orbits region and regular islands in the chaotic sea.}
\end{figure}

A typical Poincar\'e map for the parameters corresponding to the quasicrossing under investigation is presentad on Fig. \ref{f9}. Upper and lower domains filled with horizontal lines are the whispering gallery orbits. The central part corresponds to the chaotic sea with several regularity islands.
Now we can relate the classical and quantum mechanics. Figure \ref{f10} shows Husimi-Poincar\'e functions for regular and chaotic states far from the quasicrossing. The Poincar\'e surface of section is overwritten on them. As was expected, one of the states is localized\footnote{with exception of small ''tails''} on the whispering gallery orbits, and the second---in the chaotic sea. Figure \ref{f11} gives an analogous picture for one of the quasicrossing levels. It is easy to see that the Husimi-Poincar\'e function partially lies in the chaotic sea and partially---in the whispering gallery orbits region. It should be noted that in the chaotic part the function is practically zero on several stability islands and in general almost does not touch them. Transition of the function into the whispering gallery region occurs in two steps. The first to penetrate into this regular region are those parts from the chaotic sea which were initially close to it. After that the regular regions start to widen until they transform to solid line similar to that shown on the Figure \ref{f10}a. At last this solid line is joined by the remnants localized in the chaotic sea.
\section{The Conclusions}
We used a numerical method to obtain energy levels and wave functions of the quantum states in the annular billiard of the region of spectrum where chaotic effects show up. Variation of the inner disk shifting parameter in the billiard give examples of repulsion between chaotic and regular levels. Comparison of quantum functions of the Husimi-Poincar\'e quasidistributions and classical Poincar\'e surfaces of section allowed to perform detailed analysis of the correspondence between the obtained quantum states and classical orbits. For the repelling states we obtained the time evolution of the dynamical tunneling.

The demonstrated features of behavior of the quantum states near the quasicrossing point present interest for future studies. The non-trivial character of the localization of quantum states on classical objects and transitions of wave functions during the mode exchange shows that there are many problems along these directions still requiring proper solutions.

\end{document}